\documentclass[10pt,letterpaper]{article}
\usepackage{opex3}

\newcommand{\TC}{$T_C$}
\newcommand{\NN}{N$_2$}
\newcommand{\SC}{superconducting}
\newcommand{\HS}{hotspot}
\newcommand{\JC}{$J_C$}
\newcommand{\IC}{$I_C$}
\newcommand{\IB}{$I_B$}

\begin{document}

%
%

\title{High performance NbN nanowire superconducting  single photon detectors fabricated on MgO substrates}

\author{F. Marsili}

\email{francesco.marsili@epfl.ch}
\author{D. Bitauld}
\author{A. Fiore}
\address{Ecole Polytechnique F\'{e}d\'{e}rale de Lausanne
(EPFL), Institute of Photonics and Quantum Electronics (IPEQ),
Station~3,  CH-1015 Lausanne, Switzerland}

\author{A. Gaggero}
\author{F. Mattioli}
\author{R. Leoni}
\address{Istituto di Fotonica e Nanotecnologie (IFN), CNR, via del
Cineto Romano 42, 00156 Roma, Italy}

\author{M. Benkahoul}
\author{F. L\'{e}vy}
\address{ Ecole Polytechnique F\'{e}d\'{e}rale de Lausanne
(EPFL), Institute of Complex Matter Physics (IPMC), Station~3,
CH-1015 Lausanne, Switzerland }


\begin{abstract} We demonstrate high-performance nanowire
superconducting single photon detectors (SSPDs) on ultrathin NbN
films grown at a temperature compatible with monolithic integration. NbN
films ranging from 150nm to 3nm in thickness were deposited by dc magnetron sputtering on MgO substrates at $400^\circ$C. The \SC\
properties of NbN films were optimized studying the effects of deposition parameters on film properties. SSPDs were fabricated on high quality NbN
films of different thickness (7 to 3nm) deposited under optimal
conditions. Electrical and optical characterizations were performed on
the SSPDs. The highest \textit{QE} value measured at 4.2K is 20\% at 1300nm.
\end{abstract}

\ocis{(000.0000) General.} 


\bibliographystyle{osajnl}

\section{Introduction}

High counting-rate detectors capable of efficient single photon
sensing in the infrared region are needed for several applications
in different fields, including high-bandwidth interplanetary optical
communications, test of high-speed semiconductor circuits, quantum optics and quantum communications \cite{yam_natph_07}.
Nanowire superconducting  single photon detectors (SSPDs)
\cite{gol_apl_01} are interesting candidates for these
requirements. So far, high-sensitivity ultrafast SSPDs have been
fabricated only on 4nm thick NbN films grown on sapphire substrates
at high temperatures (typically $900^\circ $C) \cite{korn_ieee_05, mit_OE_06}. This
significantly limits the integration of SSPDs with advanced optical
structures (e.g. waveguides and microcavities) and read-out
electronics, typically realized on other substrates and not
compatible with these deposition temperatures. For instance, high reflectivity DBR realized on GaAs do not withstand such temperatures due to As outgassing \cite{iizuka_95}. Very recently, fabrication of SSPDs on MgO at room temperature has been reported, but their quantum efficiency (\textit{QE}) is very low \cite{miki_ieee_07}. In an effort to develop an exportable technology, we show that high performance NbN SSPDs  can be
implemented on different substrates (e.g. MgO) and at lower
deposition temperatures ($\sim400^\circ $C), which opens the way to
monolithic integration.

\section{Thin film deposition and device fabrication}

NbN films ranging from 150nm to 3nm in thickness were deposited on
epitaxial-quality single crystal MgO $<100>$ substrates by current
controlled dc magnetron sputtering from a  Nb target (5cm diameter,
99.95\% purity) in a mixture of Ar (99.9997\% purity) and \NN\ (99.999\%
purity). The background pressure was in the low $10^{-7}$ torr
range. The distance between target and substrate was fixed at the
maximum allowed by our system (85mm) to maximize film uniformity.
In order to control the thickness of ultrathin
films, the deposition rate \textit{R} was kept in the few {\AA}/s range, which
fixed the plasma current at $I_p=250$mA. Low total pressures
$P_{tot}$ (in the few mtorr range) were used, resulting in power densities of the order of 10W/cm$^2$.
The high sputtering energy allowed to promote the growth of high quality NbN films at a substrate
temperature as low as $T_S=400^{\circ}$C. Following a well
established procedure \cite{bell_jap_83,vill_ieee_84},
the \SC\ properties of NbN films were optimized studying the effects
of $P_{tot}$ and composition of reactive gas on film structural and
electrical properties. For every $P_{tot}$ the \SC\ critical
temperature \TC\ showed a maximum as a function of nitrogen partial
pressure $P_{N_2}$, which was varied within the limits determined by
the requirement to produce the NbN \SC\ $\delta$ phase
\cite{benka_SCT_04}. Decreasing $P_{tot}$ improved the quality of
NbN films: an increase in the maximum of \TC\ and a decrease in the
\SC\ transition width $\Delta$\TC\ and Residual Resistivity Ratio
(the ratio between the resistivity at 300K and at 20K,
$RRR=\rho_{300K}/\rho_{20K}$) were observed. $P_{tot}$=2.5 mtorr,
$P_{N_2}/P_{tot}=33\%$, were found to be the optimal deposition
parameters, resulting in a \TC=16.1K, with a $\Delta$\TC =60mK and
an $RRR=1$ for a 150nm thick film, which indicates that no
intergrain voids were present \cite{jones_75}. Decreasing the
thickness of films led to a slight degradation of their \SC\
properties (Fig. 1), but the thinnest film (\emph{t}=3nm) still exhibited \TC=8.6
K, $\Delta$\TC =0.9 K and $RRR=0.6$, proof of the excellent quality of
our low-temperature deposition process. The crystallinity of the best
films was characterized using X-ray diffraction: they showed a fcc,
NaCl-type crystal structure, with a lattice constant
$a_0=$4.45{\AA}, which agrees with reported results
\cite{bell_jap_83,vill_ieee_84}. From cross-sectional
transmission electron microscopy (TEM) investigation it was
determined that NbN grows on MgO substrate without any initial
amorphous layers.

\begin{figure}[htp]
\centering
\includegraphics[width=9cm]{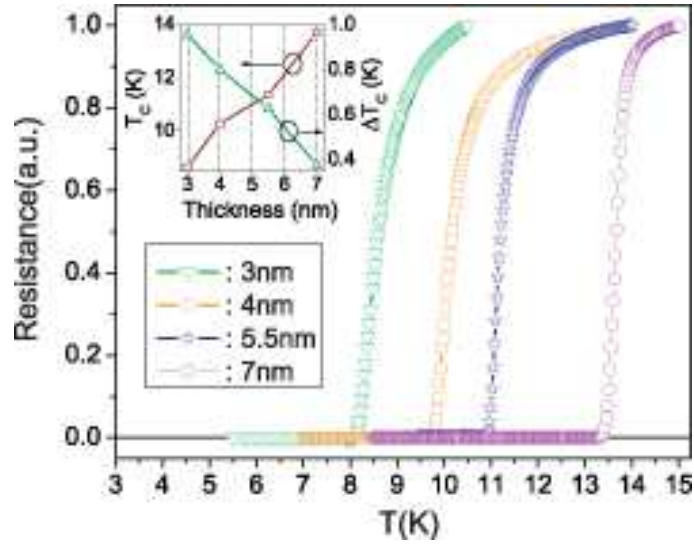}
\caption{\label{fig1}Resistance vs temperature dependence of NbN
films for four thicknesses: 7nm (circles), 5.5nm (stars), 4nm (squares)
and 3nm (triangles). The deposition conditions were:
$T_S=400^{\circ}$C, $I_p=250$mA, $P_{tot}$=2.5mtorr, 33\% \NN,
\textit{R}=3{\AA}/s. Inset: \TC\ and $\Delta$\TC\ vs. thickness (\emph{t}). \TC\ vary  from 13.7K
($\Delta$\TC=0.4K) for \emph{t}=7nm to 8.6K ($\Delta$\TC=0.9K) for \emph{t}=3nm.}
\end{figure}

SSPDs were fabricated on ultrathin NbN films deposited under optimal
conditions on MgO by a two mask process using electron beam
lithography (EBL) and reactive ion etching \cite{myjap_07}. Detectors are
5x5$\mu$m$^2$ in size, and composed of nanowires ranging from 60nm to
100nm in width (\emph{w}), folded in a meander pattern with fill factors (\emph{f})
ranging from 40\% to 60\%. The meanders are contacted through 70nm thick Au-Ti pads, patterned
as a 50$\Omega$ coplanar transmission line. The same structures were
fabricated on films of four different thicknesses (\emph{t}): 7nm, 5.5nm, 4nm
and 3nm. The thickness of NbN films was measured by AFM. \TC\ and
$\Delta$\TC\ of the patterned SSPD were found to be the same as
those of the original NbN films, which confirms that the fabrication
process does not affect their \SC\ properties.

\section{Measurements}

The uniformity in width of the nanowires was verified by extensive scanning electron microscopy (SEM). To prevent charging effects, the devices were coated by a 10nm thick layer of OsO$_4$. This conductive coating, whose grain size is less than 1nm, allowed ultra-high resolution SEM imaging on the meanders. Fig. 2 shows an SEM image of a \emph{w}=100nm, \emph{f}=40\% meander. Note that the contact pads on the two sides of the meander were also patterned with a low fill factor to reduce proximity effects during the EBL writing step. It was then possible to estimate the mean width variation as $\Delta\emph{w}\sim10nm$, which agrees with the results of electrical characterization on test structures \cite{myjap_07}.
\begin{figure}[htp]
\centering
\includegraphics[width=9cm]{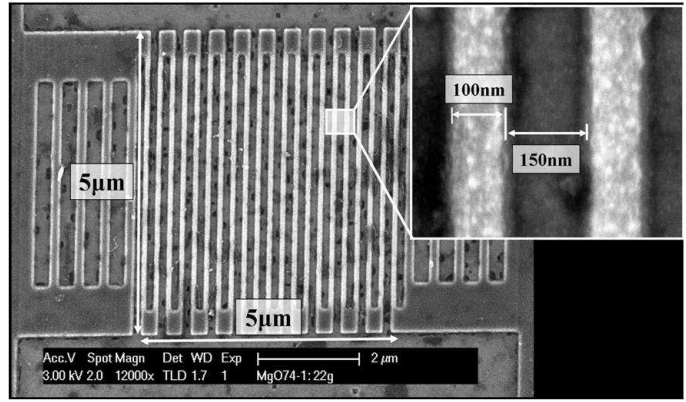}
\caption{\label{fig2} Scanning electron microscope (SEM) image of an SSPD.
The nanowire width is \emph{w}=100nm, the fill factor is \emph{f}=40\%.
The inset shows an ultra-high resolution image of two stripes. The mean width variation was estimated to be $\Delta\emph{w}\sim$10nm. }
\end{figure}

Electrical and optical characterizations have been performed on the
SSPDs. In total, 320 devices were tested, 80 for each of the four
different film thicknesses of interest. For each chip, the best
devices were first selected measuring their current-voltage (I-V)
curves inside a cryogenic probe station (Janis). Electrical contact
was realized by a cooled 50$\Omega$ microwave probe attached to a
micromanipulator, and connected by a coaxial line to the
room-temperature circuitry. The samples were DC biased using a
 low noise voltage source in series with a bias
resistor $R_B$.
\begin{figure}[htp] \centering
\includegraphics[width=9cm]{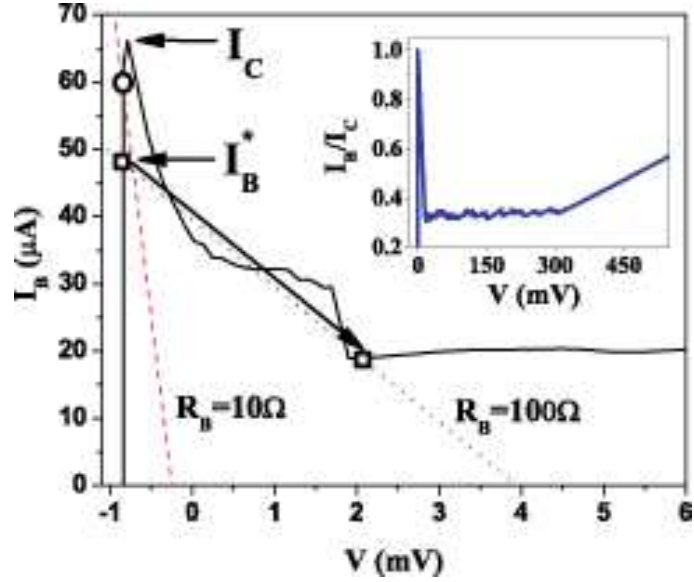}
\caption{\label{fig3}I-V curve at 4.2K of a 100nm wide, 7nm thick
meander (solid line) measured with $R_B=10\Omega$. For $R_B=10\Omega$, the DC load line (dashed)
never intersects the I-V both in the \SC\ and hotspot plateau regions, which is not the case for higher values of $R_B$ (dotted line, for $R_B=100\Omega$). The voltage offset is due to thermoelectric effects (electrical contact from room temperature to the device is realized through junctions between different metals at different temperatures, so a voltage is created due to the Seebeck effect). The inset shows the  I-V curve in a wider voltage range.}
\end{figure}
I-V curves of our devices (Fig. 3) are
typical for a \SC\ one-dimensional bridge much longer than the thermal healing
length ($\eta\sim 100$nm), at a temperature far from \TC\, and agree well with the  Skocpol-Beasley-Tinkham (SBT) hotspot model \cite{tink_jap_74}. After the current through the device exceeds the \SC\ critical current \IC, circuit-controlled relaxation oscillations are observed until, with increasing voltage, the circuit switches along the load line to the hotspot plateau. Finally, at high voltages, the extension of the hotspot approaches the total length of the nanowire and does not grow any further, so ohmic behavior is observed.
The critical current density
\JC\ was estimated from the measured value of \IC\ and the
entire geometrical cross-section of the meander. \JC\ at 4.2K varied
in the 2-4 MA/cm$^2$ range, which is a state of the art value.

The ten devices which showed the highest \JC\ in each chip were
mounted on a cryogenic dipstick and optically tested in an liquid He
bath at 4.2K.
This selection criterion relies on the fact that the most constricted segment of
a nanowire determines its \IC. Devices with a constriction (which show a low \IC) are biased well below \JC\ in most of the meander length, and thus they have a lower quantum efficiency (\textit{QE}).
Bias current was supplied through the DC port of a 10MHz-4GHz bandwidth
bias-T connected to the voltage bias circuit already described,
avoiding the latching effects associated with the current bias. The value for the bias
resistor ($R_B=10\Omega$) was chosen to attain the highest value for the bias current $I_B$ with
respect to critical current \IC\ (dashed line in Fig. 3). Increasing
$R_B$, the noise on \IB\ due to the voltage source is reduced, but
the DC load line (dotted line) become less steep. Consequently, for a given bias
current $I_B^{\ast}<I_C$ it may intersect the I-V at two points, in
the \SC\ region and in the \HS\ plateau, so that the device permanently switches from the \SC\ to the
dissipative state. The AC port of the bias-T was connected through a 4dB attenuator to the
series of two 18bB gain, 20MHz-3GHz bandwidth, low-noise amplifiers.
The amplified signal was then fed to either a 1GHz bandwidth fast
oscilloscope for time resolved measurements or a 300MHz counter for
statistical analysis.
 The devices were optically probed using  50ps wide, 25MHz repetition rate
pulses at 1.3$\mu$m wavelength from a fiber-pigtailed, gain-switched
laser diode.
 The photons were fed to the SSPDs through a  single-mode optical fiber
coupled with a 3mm focal length aspheric lens, which was placed 7cm
from the  plane of the chip in order to insure uniform illumination
of the devices. The average number of incident photons per
optical pulse was estimated to be  $\sim0.5$ with an error of 5\%.

\begin{figure}[htp]
\centering
\includegraphics[width=9cm]{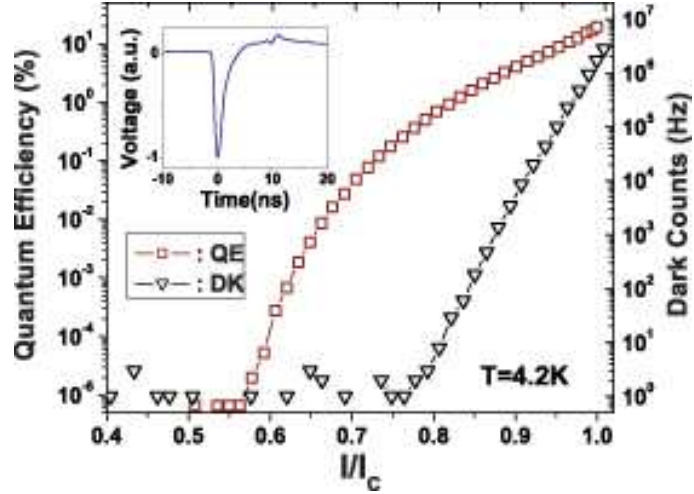}
\caption{\label{fig2}\textit{QE} (open squares) and \textit{DK} (open
triangles) as a function of the normalized bias current  for the
single photon detection regime of an optimum 5x5$\mu$m SSPD:
\textit{w}=100nm, \textit{f}=40\% t=4nm. The incident photon wavelength was 1.3$\mu$m.
Temperature was 4.2K}
\end{figure}

The dependence of the number of detector counts per second on the
average number of photons per pulse was investigated. As expected
\cite{gol_apl_02}, the dependence was linear for the photon fluxes
used in \textit{QE} measurements, proof that true single photon detection was observed.
Dark counts rate \textit{DK} was determined as the number of counts registered in one second when the SSPD
optical input was blocked.
 \textit{QE} at a certain bias current $I_B$ was calculated as: $QE=(N_c-DK)/N_{ph}$, where $N_c$ is the number of detection events registered by the counter in one second, $N_{ph}$ is the number of photons incident on the
device area in the same time and \textit{DK} is the dark counts rate at $I_B$.
The best performance was exhibited by a \emph{w}=100nm, \emph{f}=40\%, \emph{t}=4nm meander, which reaches \textit{QE}=20\% for 1.3$\mu$m wavelength light (Fig. 4) before saturation.
Using \textit{QE} and \textit{DK} from Fig. 4, noise equivalent power (\textit{NEP}, \cite{miller_apl_03}) was estimated to be of the order of $10^{-16}W/Hz^{1/2}$, which is a state of the art value at the temperature of the experiment. This is the best reported result for an SSPD detecting infrared light at 4.2K. We note that a higher \textit{QE} and a much lower \textit{DK} may be obtained by cooling the device down to 2K \cite{korn_ieee_05}. The time resolved response pulse of our best SSPD showed a full width at half maximum (FWHM) as low as 1.6ns (inset Fig. 4). No degradation in the performance was observed during the measurement period of about one month.

\section{Conclusions}

The low-temperature fabrication process has been performed also on ultrathin NbN
films deposited on GaAs. The electrical and optical measurements,
which will be reported elsewhere, are very encouraging for the
further developments of the technology on GaAs, allowing for instance monolithic integration with microcavities and waveguides. In conclusion, these results show that high performance NbN SSPDs can be realized on a different substrate and deposited at lower temperature than previously
reported, which opens the way to integration with advanced optical structures.

\section{Acknowledgments}

This work was supported by: Swiss National Foundation through the ''Professeur borsier'' and NCCR Quantum Photonics program, FP6 STREP ''SINPHONIA'' (contract number NMP4-CT-2005-16433), IP ''QAP'' (contract number 15848). The authors thank B. Dwir and H. Jotterand for technical support and useful discussion, Prof. G. Chapuis and N. Guiblin for X-ray diffraction analysis and the Interdisciplinary Centre for Electron Microscopy (CIME) for supplying TEM and SEM facilities.

\end{document}